\documentclass[10pt, a4paper]{article}
\usepackage{graphicx} 
\usepackage{subcaption} 
\usepackage{amsmath}
\usepackage{url}
\usepackage{authblk}
\usepackage{algorithm}
\usepackage{algorithmic}
\usepackage[margin=1in]{geometry}

\usepackage[compress]{cite}
\usepackage[colorlinks=true, linkcolor=blue, urlcolor=blue, citecolor=blue]{hyperref} 

\usepackage{amssymb}
\usepackage{bm}       
\usepackage{dcolumn}  

\usepackage{tikz}
\usetikzlibrary{quantikz2} 
\newtheorem{theorem}{Theorem}

\title{Block encoding of sparse matrices with a periodic diagonal structure}

\author[1]{Alessandro Andrea Zecchi\thanks{alessandroandrea.zecchi@polimi.it}}
\author[2]{Claudio Sanavio}
\author[3]{Luca Cappelli}
\author[1]{Simona Perotto}
\author[4,5]{Alessandro Roggero}
\author[2]{Sauro Succi}

\affil[1]{MOX – Department of Mathematics, Politecnico di Milano, \protect\\ Piazza L. da Vinci, 32, 20133 Milano, Italy}

\affil[2]{Fondazione Istituto Italiano di Tecnologia, \protect\\ Center for Life Nano-Neuroscience at la Sapienza, \protect\\ Viale Regina Elena 291, 00161 Roma, Italy}

\affil[3]{Dipartimento di Fisica dell’Università di Trieste, \protect\\ Via Tiepolo 11, 34131 Trieste, Italy}

\affil[4]{Physics Department, University of Trento, \protect\\ Via Sommarive 14, 38123 Trento, Italy}
\affil[5]{INFN-TIFPA Trento Institute of Fundamental Physics and Applications, \protect\\ Via Sommarive 14, 38123 Trento, Italy}

\date{\today}

\begin{document}

\maketitle

\begin{abstract}
Block encoding is a successful technique used in several powerful quantum algorithms. In this work we provide an explicit quantum circuit for block encoding a sparse matrix with a periodic diagonal structure. The proposed methodology is based on the linear combination of unitaries (LCU) framework and on an efficient unitary operator used to project the complex exponential at a frequency $\omega$ multiplied by the computational basis into its real and imaginary components. 
We demonstrate a distinct computational advantage with a $\mathcal{O}(\text{poly}(n))$ gate complexity, where $n$ is the number of qubits, in the worst-case scenario used for banded matrices, and $\mathcal{O}(n)$ when dealing with a simple diagonal matrix, compared to the exponential scaling of general-purpose methods for dense matrices. Various applications for the presented methodology are discussed in the context of solving differential problems such as the advection-diffusion-reaction (ADR) dynamics, using quantum algorithms with optimal scaling, e.g., quantum singular value transformation (QSVT). Numerical results are used to validate the analytical formulation.
\end{abstract}


\section{Introduction}
The quantum computing paradigm harnesses the features of quantum mechanics to perform computations \cite{feynman1982simulating} and is rapidly advancing toward practical applications in many research areas. Several quantum algorithms, including recent innovative developments, outperform conventional classical methods and offer significant speedup for various problems \cite{montanaro2016quantum, di_meglio_quantum_2024}. 
Moreover, many important quantum algorithms share ideas and techniques that are typically used as subroutines to perform a specific task.
For instance, the quantum Fourier transform is used for both the Shor's algorithm, which aims at finding the prime factors of an integer, and the Harrow–Hassidim–Lloyd algorithm, the latter used for finding the solution of a linear system \cite{PhysRevLett.86.1889}. 

A powerful subroutine known as block encoding emerged in the context of Hamiltonian simulation through qubitization~\cite{low2019hamiltonian}. Block encoding enables numerous complex matrix operations on quantum computers and has become a fundamental tool in developing new algorithms. This technique is widely used in quantum applications for chemistry, linear algebra, optimization, and simulating physical systems, and can be used to encode a possibly non-unitary matrix as a sub-block of a larger unitary operator \cite{Nguyen2022blockencodingdense,chakraborty_et_al:LIPIcs.ICALP.2019.33, vanapeldoorn_et_al:LIPIcs.ICALP.2019.99}.
Here we briefly recall the main mathematical aspects. Given a matrix $A\in \mathbb{C}^{2^n\times 2^n}$ we define the  unitary matrix  $U_A \in \mathbb{C}^{2^{n+m}\times 2^{n+m}}$ that provide a block encoding of $A$ with $m$ ancillary qubits, as
\begin{equation}
    U_A=
    \begin{bmatrix}
        A/\alpha & * \\
        * & *
    \end{bmatrix}.
    \label{eq:be_definition}
\end{equation}

\noindent The elements $*$ guarantee the unitarity of $U_A$ while the scaling factor $\alpha$, also called sub-normalization, ensures that $\Vert A/\alpha\Vert \leq1$ \cite{camps2024explicit} where $\Vert \cdot \Vert$ is the usual spectral matrix norm. Therefore, the scaling factor must satisfy the following inequality $\alpha\geq\Vert A \Vert$, but usually a lower scaling is favorable since the efficiency of many quantum algorithms improves as $\alpha$ reduces. Finding a block encoding scheme with $\alpha=\Vert A\Vert$ for a generic matrix $A$ remains an open problem. 

Recognizing the significance of block encoding has motivated extensive research on optimizing its construction for various classes of matrices \cite{camps2022fable, clader2023quantum, setty2025block}.
For dense matrices, i.e., with almost all elements different from zero, typically occurring when loading classical data, there has been extensive research \cite{clader2023quantum} also regarding the quantum random access memory query model \cite{PhysRevLett.100.160501}. A method, named FABLE, to generate fast approximate quantum circuits for dense matrices was proposed by \cite{9951292} and subsequently extended for unstructured sparse matrices \cite{kuklinski2024s}. Block encoding can be also used for ladder operators acting on fermionic and bosonic modes, and can be a powerful tool for simulating quantum field theories \cite{Simon2025ladderoperatorblock}.

However, currently available methods for block encoding sparse matrices often lack the ability to efficiently exploit specific structural regularities \cite{camps2024explicit}. These patterns arise frequently in engineering applications, for instance, when modeling second-order elliptic differential problems with periodic coefficients.

In this work we introduce a new block encoding method based on the LCU framework for sparse matrices with a periodic structure. In particular, we focus on sparse banded matrices and provide an explicit quantum circuit that can be easily extended to any periodic signal with a given number of frequency components. This method effectively exploits the periodicity of the matrices improving efficiency with respect to known methods and it is based on a specific unitary based on the complex exponential at a particular frequency which can be easily linked to sinusoidal functions representing its real and imaginary part.
In Section \ref{sec:I} the proposed methodology is presented and characterized for a single frequency while the extension to sparse banded matrices is considered in Section \ref{sec:be_sparse} along with the case with a more general periodic structure decomposed into different frequencies. Various applications are analyzed in Section \ref{sec:app} in the context of solving an elliptic partial differential equation (PDE) problem on a quantum computer. A paradigmatic example is provided by the classical dynamical behavior of ADR differential equation with periodically varying coefficients. The numerical results are presented in Section \ref{sec:num}. Finally, we discuss the outlooks and conclusions of our work in Section \ref{sec:conclusion}.

\section{Quantum circuit}\label{sec:I}
In this section, we describe the general idea behind our framework and provide a detailed analysis of the quantum circuit for block encoding sparse matrices with a periodic structure on the main diagonal. We follow standard conventions used in the quantum computing literature, writing quantum states in the Dirac notation $\ket{\cdot}$ and using standard quantum gates. For the binary representation of an integer $k\in\mathbb{N}$ we employ the little-endian convention with $k=k_{n-1}\cdot 2^{n-1}+\dots +k_1\cdot 2^1+k_0\cdot 2^0$ and where $k_0$ denotes the least significant bit. The qubit associated to the least significant bit appears at the bottom in quantum circuit diagrams.

\subsection{Diagonal sinusoidal matrix}
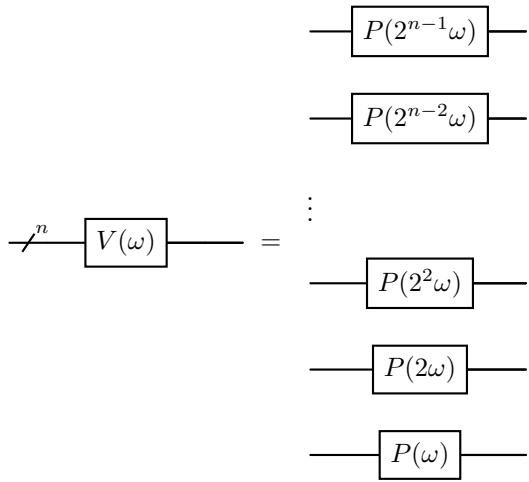
\begin{figure}[htbp]
    \centering
\begin{quantikz}
    \lstick{} & \qwbundle{n} &\gate{V(\omega)}&\qw & \qw
\end{quantikz}
=
\begin{quantikz}
    \lstick{}&\gate{P(2^{n-1}\omega)} &  \qw\\
    \lstick{}&\gate{P(2^{n-2}\omega)} &  \qw \\
    \vdots                    \\
    \lstick{}&\gate{P(2^{2}\omega)} &  \qw \\
    \lstick{}&\gate{P(2\omega)} &  \qw \\
    \lstick{}&\gate{P(\omega)} &  \qw
\end{quantikz}
    \caption{Quantum circuit to implement the diagonal unitary operator \eqref{eq:V} using $n$ phase gates.}
    \label{fig:prodP}
\end{figure}

We start by considering two diagonal matrices $C(\omega,\phi),S(\omega,\phi)\in \mathbb{R}^{N\times N}$, where $N=2^n$ with $n$ being the number of working qubits, characterized by oscillating components with frequency $\omega\in \mathbb{R}$, such that:

\begin{eqnarray}
    C(\omega,\phi)_{ij} &=& \cos((i-1)\omega+\phi)\delta_{ij} \label{eq:cos_matrix}\\
    S(\omega, \phi)_{ij} &=& \sin((i-1)\omega+\phi)\delta_{ij} \label{eq:sin_matrix},
\end{eqnarray}

\noindent where $\delta_{ij}$ is the Kronecker delta function for the matrix indices $i,j=1,\dots, N$ and $\phi\in\mathbb{R}$ is a phase term. We consider $\omega$ taking values in $[0,2\pi)$ as it does not change the overall result. In our construction, we take advantage of the fact that these two matrices are the real and imaginary components of a diagonal unitary operator  such that $(C(\omega,\phi)+\mathrm{i}S(\omega,\phi))_{ij}=e^{\mathrm{i}\left((i-1)\omega+\phi\right)}\delta_{ij}=V(\omega,\phi)$ with $\mathrm{i}$ being the imaginary unit. 
In particular, this operator can be written as
\begin{equation}
    V(\omega,\phi)=e^{\mathrm{i}\phi}\text{diag}\left(1,e^{\mathrm{i}\omega},e^{\mathrm{i}2\omega},e^{\mathrm{i}3\omega},\dots,e^{\mathrm{i}(N-1)\omega}\right),
    \label{eq:V}
\end{equation}
with $\phi$ being a global factor acting as a phase shift for the cosine and sine components. Without loss of generality, we set from now on $\phi=0$, thus writing in $V,C$ and $S$ only the dependence on $\omega$.
The unitary operator $V(\omega)$ acts on the computational basis state $\ket{k}$ as

\begin{equation}
    V(\omega)\ket{k}=e^{\mathrm{i}\omega k}\ket{k},
    \label{eq:Vact}
\end{equation}

\noindent where $k$ can be expanded in binary as $k=\sum_{q=0}^{n-1}k_q2^q$ with $k_q\in \{0,1\}$. Substituting this expression of $k$ into \eqref{eq:Vact} we can write that 

\begin{equation}
e^{\mathrm{i}\omega k}= e^{\mathrm{i}\omega\sum_{q=0}^{n-1}k_q2^q}=\prod_{q=0}^{n-1}e^{\mathrm{i}\omega k_q2^q},
\end{equation}

\noindent and since the term has been factored into a product over bits, the unitary operator $V$ can be implemented as a sequence of single-qubit rotations about the Z axis, i.e., of phase gates

\begin{equation} \label{eq:unitary_phase_V}
    V(\omega)=\prod_{q=0}^{n-1}P(2^q\omega),
\end{equation}
using the definition $P(\theta)=\text{diag}(1, e^{\mathrm{i}\theta})$.
The corresponding circuit is depicted in Figure \ref{fig:prodP}, and requires a number of single qubit phase gates equal to the number of qubits and, therefore, it can be regarded as highly efficient and attainable on a near term quantum device. Clearly, if a phase angle $\theta$ is sufficiently small, the gate's effect on the quantum state becomes negligible, allowing it to be discarded during circuit optimization \cite{Nielsen_Chuang_2010, nam_approximate_2020}. Hence, in some cases, many phase gates can be removed without significantly affecting the result, further reducing the computational cost of the operator. For instance, for small frequencies $\omega\leq2\pi/N$, we can approximate the unitary $V(\omega)$ with error at most $\epsilon$ rotating only the leading $\mathcal{O}(\log(1/\epsilon))$ qubits. This type of diagonal operator has already been studied for the Generalized Quantum Signal Processing (GQSP) framework to synthesize diagonal matrices using a Fourier decomposition into polynomials of $V(\omega)$ \cite{PRXQuantum.5.020368}. In that case, the operator was designed only for powers of the primitive $N$-th root of unity, i.e., $e^{ip\frac{2\pi}{N}}$ with $p$ being an integer, to implement convolution operators in the quantum Fourier basis. Conversely, in our case the focus is on building a block encoding of a periodic real matrix, extending subsequently the result to a linear combination of operators with different, and arbitrary, frequencies.

The interference produced by the Hadamard gates allows us to isolate the real and imaginary components of the unitary $V(\omega)$. This construction leads to the following result.

\begin{theorem} \label{thm:be}
Let $\omega \in \mathbb{R}$ and let $V(\omega)=V(\omega,0)$ be the diagonal unitary operator defined in Equation ~\eqref{eq:V}. The unitary $U_{C(\omega)}$ acting on $n+1$ qubits, defined as 
\begin{equation}
U_{C(\omega)}=( H\otimes I)(I\otimes V(\omega)^\dagger ) ( \ket{0}\bra{0} \otimes V(2\omega) + \ket{1}\bra{1} \otimes I) ( H\otimes I ),
\label{eq:UA_general}
\end{equation}
is an exact block encoding of the diagonal matrix $C(\omega)=C(\omega,0)$ defined in Equation \eqref{eq:cos_matrix} with sub-normalization factor $\alpha=1$ and $m=1$ ancilla qubit.
\end{theorem}

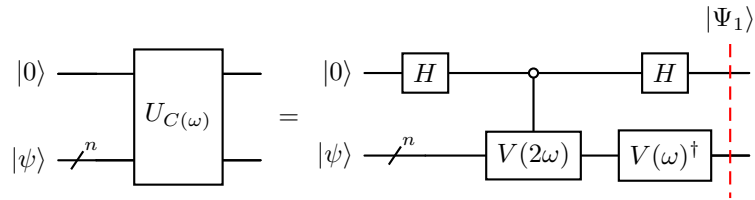
\begin{figure}[htbp]
    \centering
\begin{quantikz}
    \lstick{$\ket{0}$} & \qw & \gate[2]{U_{C(\omega)}} & \qw \\
    \lstick{$\ket{\psi}$} & \qwbundle{n} & \qw & \qw
\end{quantikz}
=
\begin{quantikz}
    \lstick{$\ket{0}$}&\gate{H} &\octrl{1} & \gate{H}\slice{$\ket{\Psi_1}$} & \qw \\
    \lstick{$\ket{\psi}$}&\qwbundle{n}&\gate{V(2\omega)} &\gate{V(\omega)^\dagger}&\qw
\end{quantikz}
    \caption{Block encoding circuit of the diagonal matrix $C$ defined in Equation \eqref{eq:cos_matrix}.}
    \label{fig:circuit}
\end{figure}

The proof follows from the direct application of the unitary operators and the identity $V(\omega)^\dagger=V(-\omega)$. If we consider an ancilla qubit initialized in $|0\rangle$ with a working register of size $n$ having an arbitrary initial state $\ket{\psi}$, as depicted in Figure \ref{fig:circuit}, the resulting overall state corresponds to the superposition  
\begin{eqnarray} \label{eq:resultbe}
    \ket{\Psi_1}=|0\rangle C(\omega)|\psi\rangle+\mathrm{i}|1\rangle S(\omega)|\psi\rangle.
\end{eqnarray}

\noindent The state $\ket{\Psi_1}$, depending on the result of the measurement performed on the ancilla qubit, collapses into either the state $C(\omega)|\psi\rangle$ or the state $S(\omega)|\psi\rangle$ on the working qubits up to a global phase.

\noindent We remark that by recovering the dependency on the phase factor $\phi$ it is possible to retrieve the more general expressions of Equations \eqref{eq:cos_matrix} and \eqref{eq:sin_matrix}. In addition, it is sufficient to add a Pauli-$Y$ gate to the ancilla qubit to obtain
\begin{eqnarray} \label{eq:resultbe2}
    (Y\otimes I)\ket{\Psi_1}=Y(|0\rangle C(\omega)|\psi\rangle+\mathrm{i}|1\rangle S(\omega)|\psi\rangle)=\ket{0}S(\omega)\ket{\psi}+\mathrm{i}\ket{1}C(\omega)\ket{\psi},
\end{eqnarray}
thus achieving a block encoding of the \textit{sine} matrix $S(\omega)$ associated with the $\ket{0}$ state of the ancilla qubit.

The probability of the two outcomes of Equation \eqref{eq:resultbe} depends on the initial state $|\psi\rangle$. If we write the state as $|\psi\rangle = \sum_{k=0}^{N-1}c_k|k\rangle$, with $\sum_{k=0}^{N-1}|c_k|^2=1$, then the probability of measuring the state $|0\rangle$ is $p_0=\sum_{k=0}^{N-1}|c_k|^2\cos^2(k\omega)$, while the probability of measuring $|1\rangle$ is $p_1 = 1-p_0 = \sum_{k=0}^{N-1}|c_k|^2\sin^2(k\omega)$.

\begin{figure}[htbp]
\centering
\includegraphics[width=0.55\linewidth]{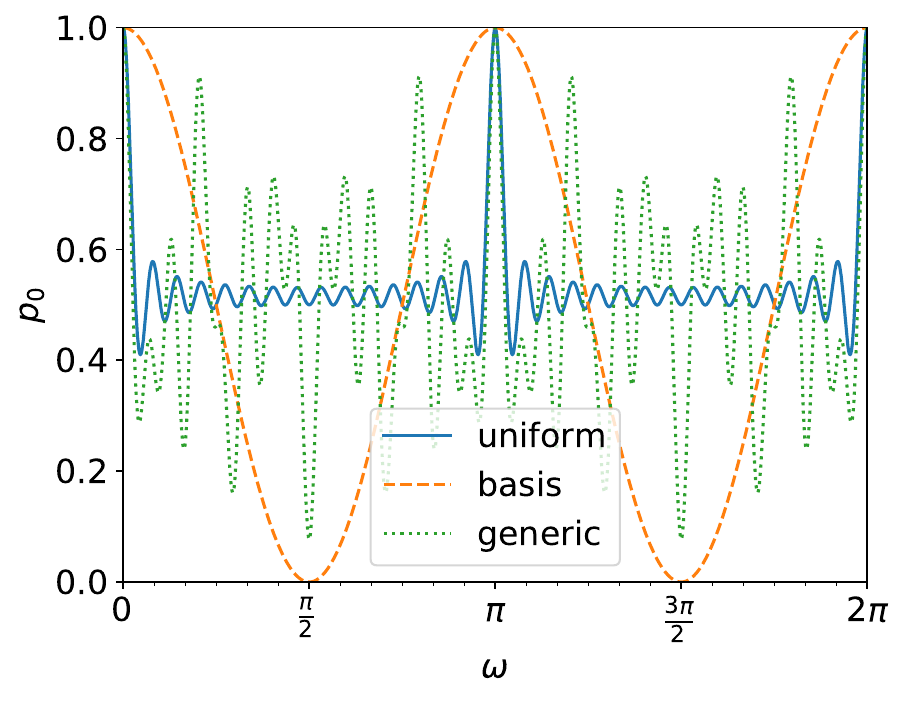}
\caption{Probability $p_0$ of successfully applying the proposed block encoding for different values of $\omega$ and different choices of the initial state $\psi$, the uniform distribution (solid), the computational basis case (dashed) and a generic distribution (dotted). The quantum register of $\psi$ consists of 4 qubits.}
\label{fig:plot_compare}
\end{figure}

Employing the trigonometric identity $\cos(x)^2=1/2(1+\cos(2x))$ it is also possible to write $p_0=1/2+1/2\sum_{k=0}^{N-1}|c_k|^2\cos(2k\omega)$ which shows that the probability oscillates around an \textit{average} value of $1/2$ and depends both on the initial coefficients $c_k$ and on the frequency $\omega$. The probability $p_0$ is shown in Figure \ref{fig:plot_compare} for an initial uniform distribution (or superposition)  with all equal coefficients, while for a single computational basis state $k$ it oscillates exactly as $p_0=\cos(k\omega)^2$. Instead, for a generic distribution the probability oscillates between 0 and 1 with an interference-like pattern.

\begin{figure}[htbp]
    \centering
    \begin{quantikz}
        \lstick{$\ket{0}$} & \qw & \gate[2]{U_{C(\omega)}} & \qw \\
        \lstick{$\ket{\psi}$} & \qwbundle{n} & \qw & \qw
    \end{quantikz}
    =
    \begin{quantikz}
        \lstick{$\ket{0}$} & \gate{H} & \ctrl{1} & \qw & \ctrl{1} & \gate{H} & \qw \\
        \lstick{$\ket{\psi}$} & \qwbundle{n} & \targ{} & \gate{V(\omega)} & \targ{} & \qw & \qw
    \end{quantikz}
    \caption{Alternative block encoding circuit of the diagonal matrix $C$ defined in Equation \eqref{eq:cos_matrix}. The CNOT gates have one control and $n$ targets.}
    \label{fig:circuit_new}
\end{figure}
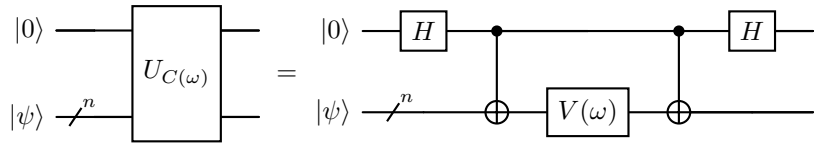

We point out that the proposed circuit for block encoding a diagonal non unitary matrix (see Figure \ref{fig:circuit}) has the advantage of requiring less controlled multi-qubit operations with respect to other schemes designed for more general matrices\cite{10.1145/3718348}. The circuit in Figure~\ref{fig:circuit} can be implemented straightforwardly using $3n$ phase gates and $2n$ CNOT gates~\cite{barenco_1995}. An even simpler implementation requiring only $n$ phase gates instead can be obtained using the implementation shown in Figure~\ref{fig:circuit_new}.

\subsection{Shift permutation matrices}

The left $L$ and right $R$ shift permutation matrices are defined as follows
\begin{equation}
L = \begin{bmatrix}
0 & 0 & \cdots & \cdots & 1 \\
1 & 0 & 0 & \cdots & 0 \\
\vdots & 1 & \ddots & \ddots & \vdots \\
\vdots & \ddots & \ddots & \ddots & \vdots \\
0 & 0 & \cdots & 1 & 0
\end{bmatrix}, 
\quad
R = \begin{bmatrix}
0 & 1 & \cdots & \cdots & 0 \\
0 & 0 & 1 & \ddots & 0 \\
\vdots & 0 & \ddots & \ddots & \vdots \\
\vdots & \vdots & \ddots & \ddots & 1 \\
1 & 0 & \cdots & \cdots & 0
\end{bmatrix}.
\end{equation}

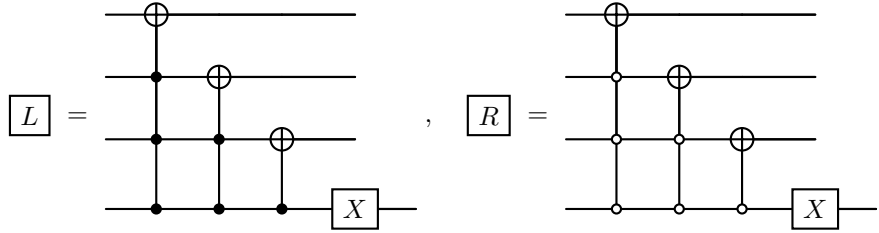
\begin{figure}[htbp]
    \centering
    \begin{quantikz}
    \gate{L}
    \end{quantikz} = 
\begin{quantikz}
      \qw & \targ{}    & \qw       & \qw      & \qw \\
      \qw & \ctrl{-1}  & \targ{}   & \qw      & \qw \\
      \qw & \ctrl{-2}  & \ctrl{-1}   &   \targ{}   & \qw \\
      \qw & \ctrl{-3}  & \ctrl{-2}      & \ctrl{-1} & \gate{X} & \qw
    \end{quantikz},\quad
    \begin{quantikz}
    \gate{R}
    \end{quantikz} = 
\begin{quantikz}
      \qw & \targ{}    & \qw       & \qw      & \qw \\
      \qw & \octrl{-1}  & \targ{}   & \qw      & \qw \\
      \qw & \octrl{-2}  & \octrl{-1}   &   \targ{}   & \qw \\
      \qw & \octrl{-3}  & \octrl{-2}      & \octrl{-1} & \gate{X} & \qw
    \end{quantikz}
    \caption{Shift circuits for a quantum register of size 4.}
    \label{fig:shift_circuit}
\end{figure}

These operators correspond to the addition and subtraction arithmetic operations modulo $N$ and can be used to apply a shift to a diagonal matrix. The shift operators can be constructed in a straightforward way following the circuit in Figure ~\ref{fig:shift_circuit} using a sequence of multi-qubit controlled Toffoli gates. In this case the depth is quadratic in $n$ and  the number of Toffoli gates scales as $\mathcal{O}(n^2)$, using a single \textit{dirty} (without requiring re-initialization)  ancilla qubit~\cite{zindorf_efficient_2025}. Further savings can be achieved using the approach applied to the quantum adder in \cite{Gidney2018halvingcostof} which requires $n-1$ clean ancilla qubits but has a Toffoli cost of only $\mathcal{O}(n)$. 

By iteratively applying $s$ times one of these operators, i.e., $R^s$ or $L^s$, further shifts to right or the left, respectively, can be achieved, while appending one of these operators to the working register of the circuit defined in Figure \ref{fig:circuit} is an effective strategy to shift the periodic structures of Equations \eqref{eq:cos_matrix} and \eqref{eq:sin_matrix}. For instance, the result of applying the right shift operator to the matrix $C(\omega)$ is
\begin{equation}
    RC(\omega)=\begin{bmatrix}
0 & \cos(\omega) & \cdots & \cdots & 0 \\
0 & 0 & \cos(2\omega) & \ddots & 0 \\
\vdots & 0 & \ddots & \ddots & \vdots \\
\vdots & \vdots & \ddots & \ddots & \cos((N-1)\omega) \\
1 & 0 & \cdots & \cdots & 0
\end{bmatrix}.
\end{equation}

\section{Block encoding of sparse matrices with a periodic structure}\label{sec:be_sparse}

We consider a sparse banded matrix $M\in \mathbb{R}^{N\times N}$, again with $N=2^n$, whose off-diagonal configuration is Toeplitz with wrap-around and with an arbitrary structure on the main diagonal, namely

\begin{equation}
    M = 		\begin{bmatrix}
m_{11} & \alpha_2 & 0 & \cdots & \alpha_1 \\
\alpha_1 & m_{22} & \ddots & \ddots & 0 \\
0 & \alpha_1 & \ddots & \alpha_2 & \vdots \\
\vdots & \ddots & \ddots & \ddots & \alpha_2 \\
\alpha_2 & 0 & \cdots & \alpha_1 & m_{NN}
\end{bmatrix}.
\label{eq:matM}
\end{equation}

This kind of matrix can arise when dealing with a directed cyclic graph with different weights for each vertex, or when dealing with the algebraic form resulting from the finite discretizations of partial differential equations with periodic boundary conditions \cite{doi:10.1137/1.9780898717839, quarteroni2010numerical}.

It is well known that by using the Linear Combinations of Unitary operations (LCU) framework, originally developed in \cite{10.5555/2481569.2481570}, it is possible to decompose any operator $O$ into a sum of $J$ unitary matrices $U_j$, each one weighted by a real positive coefficient $\alpha_j$, for $j=0,1,\dots,J-1$, as

\begin{equation}
    O =\sum_{j=0}^{J-1}\alpha_jU_j.
\end{equation}

In particular, the linear combination is achieved using a unitary state preparation operator $PREP$, and its inverse $PREP^\dagger $.
More precisely, we can define the state preparation operator, acting onto a data register of $\lceil\log_2(J)\rceil$ qubits as
\begin{equation}
    PREP\ket{0}=\sum_{j=0}^{J-1}\sqrt{\frac{\alpha_j}{\alpha}}\ket{j},
\end{equation}
with $\alpha=\sum_{j=0}^{J-1} \alpha_j $ being a proper normalization coefficient. In order to obtain the full operator $O$ we also need a $SELECT$ unitary defined as
\begin{equation}
SELECT = \sum_{j=0}^{J-1} \rvert j\rangle\langle j\lvert\otimes U_j\;,
\end{equation}
so that we can write a rescaled version of our operator as follows
\begin{equation}
\frac{O}{\alpha} = \langle 0\lvert \left(PREP\right)^\dagger\left( SELECT \right)\left(PREP\right) \rvert 0\rangle\;.
\end{equation}

If the matrix $M$ has a periodic structure on the main diagonal, we can employ the matrix $C(\omega)$ defined in Equation \eqref{eq:cos_matrix}, and the left and right shift permutation matrices, to construct the following LCU, involving four positive real coefficients

\begin{equation}
    M=\alpha_0C(\omega)+\alpha_1L+\alpha_2R+\alpha_3I.
    \label{eq:M_dec_I}
\end{equation}

\begin{figure}[htbp]
     \centering
\begin{quantikz}
    \lstick{$\ket{d}$} & \qw & \gate[3]{U_{M(\omega)}} & \qw \\
    \lstick{$\ket{0}$} & \qw & \qw & \qw \\
    \lstick{$\ket{\psi}$} & \qwbundle{n} & \qw & \qw
\end{quantikz}
=
    \begin{quantikz}[row sep=0.5cm, column sep=0.5cm]
        \lstick{\ket{d_1}=\ket{0}} & \qw & \qw & \gate[2]{PREP} & \qw & \octrl{2} & \qw & \octrl{3} & \qw & \ctrl{3} & \qw & \gate[2]{PREP^\dagger}& \qw\\
        \lstick{\ket{d_0}=\ket{0}} & \qw & \qw & \qw             & \qw & \octrl{1} & \qw & \ctrl{2} & \qw & \octrl{2} & \qw & \qw& \qw\\
        \lstick{\ket{0}} & \qw & \qw & \qw             & \qw & \gate[2]{U_{C(\omega)}} & \qw & \qw & \qw & \qw & \qw & \qw & \qw\\
        \lstick{\ket{\psi}} & \qwbundle{n} & \qw & \qw & \qw & \qw  & \qw & \gate{L} & \qw & \gate{R} & \qw & \qw& \qw\\
    \end{quantikz}
    \caption{Block encoding of a sparse banded matrix with a periodic structure employing LCU.}
    \label{fig:lc_circuit}
\end{figure}
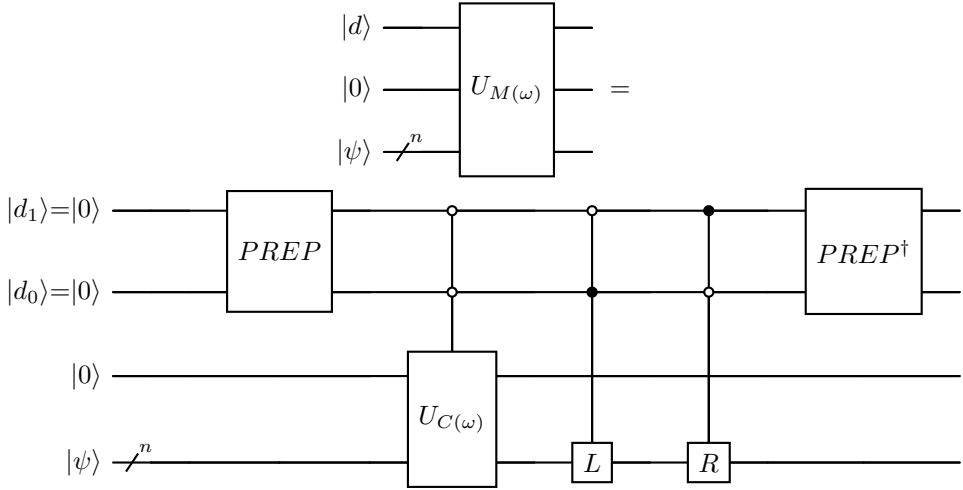

In this case the resulting matrix is defined as in Equation \eqref{eq:matM} but with diagonal entries given by 
\begin{equation}
    m_{ii}=\alpha_0\cos((i-1)\omega+\phi)+\alpha_3, \quad i=1,\dots ,N.
\end{equation} 
\noindent The quantum circuit depicted in Figure \ref{fig:lc_circuit}, is a block encoding of $M$ using three ancillary qubits: one for the block encoding of $C(\omega)$ (see Theorem  \ref{thm:be}), and two for encoding the LCU coefficients into an appropriate quantum state. Post-selecting the ancillary qubits in the state $\ket{000}$ implements the desired matrix transformation.

This quantum circuit can be considered efficient as the gate complexity is $\mathcal{O}(n^2)$ as seen from the analysis above where we have fully characterized its various components. The dominant cost is the implementation of the $SELECT$ circuit which requires $J$ multiply-controlled operations. The controlled version of $U_{C(\omega)}$ can be obtained by controlling the $V(\omega)$ unitary in Figure~\ref{fig:circuit_new} from the two ancilla qubits. This can be done easily by adding two Toffoli with target an additional clean ancilla, $n$ additional rotations and $2n$ CNOT gates controlled on the added ancilla. When using the unitary incrementer circuit from \cite{Gidney2018halvingcostof} to implement the shift operators $L$ and $R$, their controlled version can be implemented using a constant number of additional gates. Finally, the Toffoli cost of implementing the sequence controls needed for the $SELECT$ operation can be brought down to only $\mathcal{O}(J)$ using the unary iteration scheme proposed in \cite{PhysRevX.8.041015}, which organizes these controls into a ladder structure to optimize circuit depth. 
We also remark that for this block encoding scheme it is also possible to modify the quantum circuit to zero out the $\alpha_1$ in the $(1,N)$-th entry of $M$ and the $\alpha_2$ in the $(N,1)$-th entry of $M$, obtaining a tridiagonal matrix following the procedure explained in \cite{camps2024explicit}.

We provide for sake of concreteness a simple example with $n=2$ working qubits, $\omega=\pi/2$ and using $PREP=H\otimes H$, therefore with $\alpha_j=1/4, \quad \forall j$ for which the resulting matrix is simply given by

\begin{equation}
    M = 	\frac{1}{4}	\begin{bmatrix}
\cos(0)+1 & 1 & 0 & 1 \\
1 & \cos(\pi/2)+1 & 1 &  0 \\
0 & 1 & \cos(\pi)+1 & 1 \\
1 & 0 & 1 & \cos(3 \pi /2)+1 \\
\end{bmatrix},
\label{eq:Mexample}
\end{equation}

with the full circuit implementation given by Figure \ref{fig:lc_circuit}.

\subsection{Fourier decomposition: periodic structure with more than one frequency}

If we consider a generic signal it is possible to employ Fourier analysis to decompose it into its basic components. Employing this tool it is possible to express the original signal as a linear combination of periodic components at different frequencies. The same concept can be applied for the elements of a generic diagonal matrix. In particular, using the LCU framework it is also possible to extend the previous results to include matrices defined as in \eqref{eq:matM}, that exhibit a periodic structure with multiple frequencies.
To introduce this concept for diagonal matrices, let us consider only two different frequencies $\omega_1,\omega_2$ and without any phase factor term. In this case, the general matrix with cosines on the diagonal can be written using a weighted combination with two coefficients. More generally, if we consider a signal consisting of $p$ cosines with different frequencies an LCU with $p$ terms corresponds to
\begin{equation}
    \sum_{j=0}^{p-1}\alpha_jU_{C(\omega_j)},
\end{equation}
with a complexity of $\mathcal{O}(p \log N)$. This result is a generalization of the GQSP framework \cite{PRXQuantum.5.020368} since it can express the frequency $\omega$ not necessarily as a root of unity $2\pi/N$. We can reformulate this result including also sine waves. In particular, given a generic $T-$periodic real function expressed as a trigonometric series with appropriate coefficients
\begin{equation}
    f(x)=\frac{a_0}{2}+\sum_{n=1}^\infty\left[a_n\cos\left(\frac{2\pi}{T}nx\right)+b_n\sin\left(\frac{2\pi}{T}nx\right)\right],
\end{equation}
it is possible to implement a block encoding of the $N\times N$ diagonal matrix $\mathcal{A}$ such that
\begin{equation}
    \mathcal{A}\ket{x}=f(x)\ket{x},
\end{equation}
using an LCU employing block encodings of the \textit{cosine} and \textit{sine} matrix defined in Equations \eqref{eq:cos_matrix} and \eqref{eq:sin_matrix}.
The proposed methodology can be more efficient than the one used in GQSP which relies on a signal decomposition into a $d-$degree polynomial of $V(\omega)$ requiring $\mathcal{O}(d \log N)$ 1 and 2 qubits gates. The two methods are equivalent when $\omega$ can be written as an integer multiplied by the root of unity $2\pi/N$. However, in the more general case, the order $d$ needed for the GQSP-based approach can be much larger than the order $p$ required by our method. A simple example is for a matrix proportional to $C(\pi N/(N+1))$ requiring only $p=1$ for our method, which then costs $\mathcal{O}(\log(N))$, but for GQSP the cost becomes $\mathcal{O}(N\log(N))$ due to the need to use a $d=N$ polynomial. It remains to be explored whether this approach can be generalized using matrix access oracles \cite{camps2024explicit} and whether that would result in a computational advantage. Specifically, an open research question is whether it is possible to design a oracle to pick between various frequencies, such that the final block encoded operator represents the sum of different components.

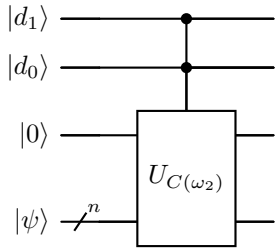
\begin{figure}[htbp]
    \centering
    \begin{quantikz}[row sep=0.5cm, column sep=0.5cm]
        \lstick{\ket{d_1}} &    \qw & \ctrl{2} & \qw  \\
        \lstick{\ket{d_0}} &   \qw & \ctrl{1} & \qw  \qw\\
        \lstick{\ket{0}} &  \qw & \gate[2]{U_{C(\omega_2)}} & \qw \\
        \lstick{\ket{\psi}} & \qwbundle{n} & \qw  & \qw \\
    \end{quantikz}
    \caption{Controlled operation to include another frequency. An additional ancillary qubit for the block encoding of $U_{C(\omega_2)}$ is employed.}
    \label{fig:lc_additional}
\end{figure}

Let us now discuss the case of periodic matrices of Equation \eqref{eq:matM}. Considering again only two frequencies, using four positive real coefficients for the LCU it is possible to write the overall matrix $M$ expanded as
\begin{equation}
    M=\alpha_0C(\omega_1)+\alpha_1L+\alpha_2R+\alpha_3C(\omega_2),
\end{equation}
\noindent with corresponding diagonal entries
\begin{equation}
    m_{ii}=\alpha_0\cos((i-1)\omega_1)+\alpha_3\cos((i-1))\omega_2), \quad i=1,\dots,N.
\end{equation}
In Figure \ref{fig:lc_additional} it is shown the additional operator required for this case, which needs to be included in the previous LCU circuit between the $PREP$ and $PREP^\dagger$ unitaries.

\section{Applications} \label{sec:app}

The simulation of classical systems, in the form of partial differential equations, is of major interest for many engineering and scientific applications. Investigating the potential computational advantage enabled by quantum algorithms in addressing this task is of substantial importance. In this section we apply the proposed methodology to different model problems.

\subsection{Elliptic problem with periodically varying reaction}
As a first example, we consider the following one-dimensional elliptic problem with periodic boundary conditions
\begin{equation}\label{eq:ellipticproblem}
\begin{cases}
-Du''+au=f \quad x\in(0,1)\\
u(0)=u(1)    
\end{cases}
\end{equation}
with a constant diffusion coefficient $D=1$ and a positive periodically varying reaction term $a(x)=a_0+\cos(\omega x)$ with $a_0>1$. We emphasize that the reaction term does not need to be periodic with period equal to the domain length. This formulation can be used to find the ground state of a particle in a periodic potential. For simplicity,  we consider also a constant real coefficient for the source term $f=1$. Using a centered finite difference scheme to approximate the second derivative, on a uniform discretization of spacing $h=1/N$ inside $(0,1)$, the problem can be written as a linear system of equations of dimension $N$ 
\begin{equation}
    A\mathbf{u}=\mathbf{f},
\end{equation} 
where $\mathbf{u}=(u_1,u_2,\cdots,u_N)^T$ is the vector of unknowns, $\mathbf{f}=(1,1,\cdots,1)^T$ is the right-hand side vector, and $A$ is the tridiagonal banded matrix
\begin{equation}
    A = 		1/h^2\begin{bmatrix}
2+a_1h^2 & -1 & 0 & \cdots & -1 \\
-1 & 2+a_2h^2 & \ddots & \ddots & 0 \\
0 & -1 & \ddots & -1 & \vdots \\
\vdots & \ddots & \ddots & \ddots & -1 \\
-1 & 0 & \cdots & -1 &2+a_Nh^2
\end{bmatrix}.
\end{equation}
 We use the notation $u_j=u(x_j), a_j=a(x_j)$ for the nodes $x_j=jh-h$, with $j=1,2,\dots,N$ and the fact that $u(x_1)=u(x_{N+1})$. This symmetric matrix can be block encoded using the circuit of Figure \ref{fig:lc_circuit} and a negative sign to the left and right shift permutation matrices. It is possible to solve this system by employing the QSVT algorithm for matrix inversion\cite{martyn_grand_2021}. More in detail, QSVT requires access to the block encoding of the linear system matrix and to prepare an initial state proportional to $\mathbf{f}$. The application of the matrix $A^{-1}$ is achieved through a spectral transformation to the singular values of $A$ using a polynomial approximation of the function $f(x)=\frac{1}{x}$.

\subsection{ADR system with periodically varying reaction term}

We consider here a prototypical case being the advection-diffusion-reaction system and discuss the feasibility of a quantum simulation method applying the proposed block encoding scheme. The ADR model can be used, for instance, in morphogenesis to describe how biological forms, patterns, and tissue shapes arise through the interaction of chemical signaling (reaction), molecular spreading (diffusion), and transport by fluid flow (advection) \cite{10.1093/oso/9780198503989.001.0001, PhysRevE.83.016702, CHOTIBUT2017500}. 

We simplify our discussion to a one-dimensional setting of length $L$, and define the problem as

\begin{equation}
\begin{cases}
    \displaystyle\frac{\partial \psi }{\partial t}=\frac{\partial }{\partial x}\left ( D\frac{\partial \psi}{\partial x}\right)-\frac{\partial(c\psi)}{\partial x}-a\psi   & x\in (0,L), t>0\\
    \psi(x,0)=g(x) & x\in (0,L)\\
    \psi(0,t)=\psi(L,t) & t>0
\end{cases}
\label{eq:ADRprob}
\end{equation}

\noindent where $\psi(x,t)$ is the unknown scalar function with initial condition $\psi(x,0)=g(x)$ and periodic boundary conditions are set. We consider a constant diffusion coefficient $D \in \mathbb{R}^+$, a constant velocity coefficient $c \in \mathbb{R}$ and $a:(0,L)\rightarrow\mathbb{R}^+, \quad a(x)=a_0+\cos(\omega x)$ is a positive reaction coefficient with a periodic component oscillating with (spatial) period $2\pi/\omega$. This is an extension of the analysis provided in \cite{10899872}, which considered only constant coefficients. This model can also be used for the Fokker-Planck equation, where the statistical distribution of particle velocities evolves under the combined effects of drag and Brownian diffusion \cite{risken1989fokker}. In addition, if one sets $c=0$ and uses quantum imaginary time evolution \cite{motta2020determining}, this equation describes the unitary dynamics of a wave-packet in a one-dimensional domain under the influence of an external periodic potential.

Employing a centered finite difference scheme to approximate the derivatives with respect to the variable $x$ on a uniform distribution of $N$ spatial nodes the problem can be written in an algebraic form as follows

\begin{equation}
\begin{cases}
\displaystyle\frac{d\bm{\psi}(t)}{dt} = M \bm{\psi}(t), \quad t>0 \\
\bm{\psi}(0) = \mathbf{g}(x)
\end{cases}
\label{eq:discr}
\end{equation}
where the matrix $M$ is a banded matrix with a periodic structure on the main diagonal
\begin{equation}
   M_{ij} =
\begin{cases}
\displaystyle -\frac{2D}{\Delta x^2} - a_0-\cos(\omega x_i) & \text{if } j = i \\
\noalign{\vskip9pt} 
\displaystyle \frac{D}{\Delta x^2} - \frac{c}{2\Delta x} & \text{if (} j = i+1\text{) or } (i=N \text{ and } j=1) \\
\noalign{\vskip9pt} 
\displaystyle \frac{D}{\Delta x^2} + \frac{c}{2\Delta x} & \text{if (} j = i-1  \text{) or } (i=1 \text{ and } j=N) \\
\noalign{\vskip9pt} 
0 & \text{otherwise}.
\end{cases}
\label{eq:matrixdiscper}
\end{equation}

It is possible to apply the forward Euler method to numerically solve \eqref{eq:discr} and obtain 

\begin{equation}
\begin{cases}
\bm{\psi}(t+\Delta t) = (I+\Delta tM) \bm{\psi}(t) = A \bm \psi(t) \\
\bm{\psi}(0) = \mathbf{g}(x)
\end{cases}
\label{eq:feuler}
\end{equation}

where here the matrix $A$ is, again, a banded matrix with a periodic structure on the main diagonal. In order to guarantee stability of the Forward Euler scheme, the time step $\Delta t$ must satisfy a CFL condition depending on the diffusion, advection, and reaction coefficients. Both the matrix $M$ and $A$ can be decomposed as in equation \eqref{eq:M_dec_I}.

\subsection{ADR system with periodically varying velocity}
As a final case we model again the ADR problem \eqref{eq:ADRprob} in the one-dimensional domain but now with a periodically varying velocity $c=\sin(\omega x)$ term and constant diffusion and reaction terms $D,a\in \mathbb{R}^+$. In this setting the advection term can be written as

\begin{equation}
    \frac{\partial(c\psi)}{\partial x}=\frac{\partial c}{\partial x}\psi +c\frac{\partial\psi}{\partial x}=\omega\cos(\omega x)\psi+\sin(\omega x)\frac{\partial\psi}{\partial x}.
\end{equation}
Therefore, the \textit{effective} reaction term is $a+\omega\cos(\omega x)$ which can be treated as in the previous case. By applying again a centered finite difference scheme and the forward Euler method as in \eqref{eq:feuler}, the resulting matrix $A$ is

\begin{equation}
   A_{ij} =
\begin{cases}
\displaystyle -\frac{2D}{\Delta x^2} - a+\omega \cos(\omega x_i) & \text{if } j = i \\
\noalign{\vskip9pt} 
\displaystyle \frac{D}{\Delta x^2} - \frac{\sin(\omega x_i)}{2\Delta x} & \text{if (} j = i+1\text{) or } (i=N \text{ and } j=1) \\
\noalign{\vskip9pt} 
\displaystyle \frac{D}{\Delta x^2} + \frac{\sin(\omega x_i)}{2\Delta x} & \text{if (} j = i-1  \text{) or } (i=1 \text{ and } j=N) \\
\noalign{\vskip9pt} 
0 & \text{otherwise},
\end{cases}
\label{eq:matrixdiscperc}
\end{equation}
and can be written as an LCU
\begin{equation}
A=\alpha_0C(\omega)+\alpha_1L+\alpha_2LS(\omega)+\alpha_3R+\alpha_4(-RS(\omega))+\alpha_5(-I),
\end{equation}
where the coefficients $\{\alpha\}_{j=0}^5$ are real numbers depending on the specific values of $D$ and $a$.

\section{Numerical simulations} \label{sec:num}
\begin{figure}[htbp]
    \centering
    \includegraphics[width=0.55\linewidth]{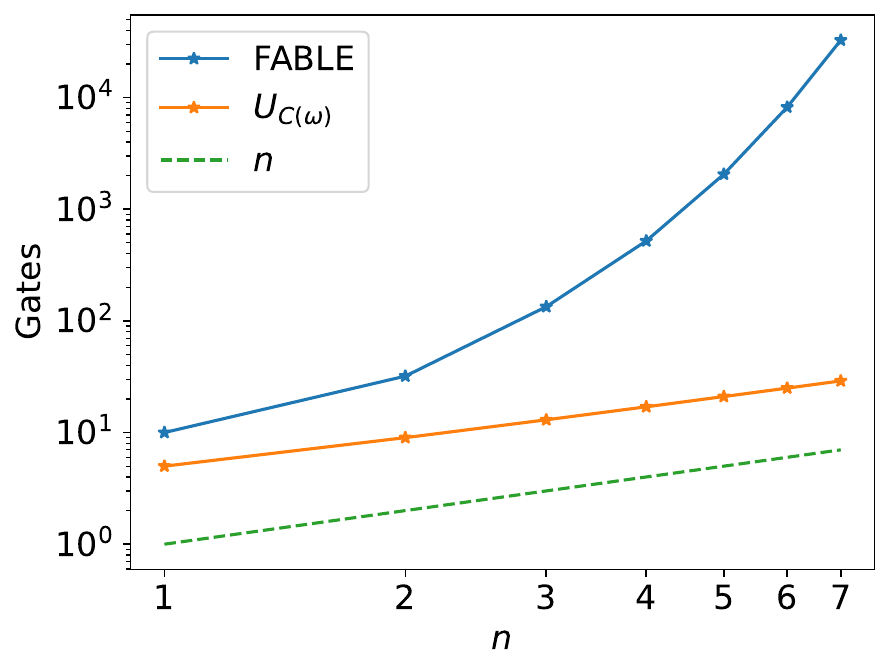}
    \caption{Number of operations with respect to the number of qubits $n$ for the block encoding of the diagonal matrix $C(\omega)$ using the FABLE library and the block encoding scheme proposed in this work $U_{C(\omega)}$ for $\omega=2$. The circuit are constructed with Qiskit and transpiled to a generic quantum hardware counting single and two-qubits gates. The scale is logarithmic for both axes.}
    \label{fig:compareFABLE}
\end{figure}

In this section we present numerical results for the proposed methodology and the previously mentioned applications. Given the periodic matrix $C(\omega)$ defined in Equation \eqref{eq:cos_matrix} we compared the proposed methodology to the FABLE method to analyze different block encoding strategies.
As shown in Figure \ref{fig:compareFABLE}, the block encoding of this work $U_{C(\omega)}$ has a linear complexity compared to the more general FABLE method which scales exponentially with respect to the number of qubits $n$. The linear scaling is replaced for a banded matrix, defined by the LCU of Equation \eqref{eq:M_dec_I}, by a either a linear or quadratic one depending on the implementation of the $L$ and $R$ matrices.

\begin{figure}[htbp]
    \centering
    \begin{subfigure}[b]{0.49\textwidth}
        \centering
        \includegraphics[width=\textwidth]{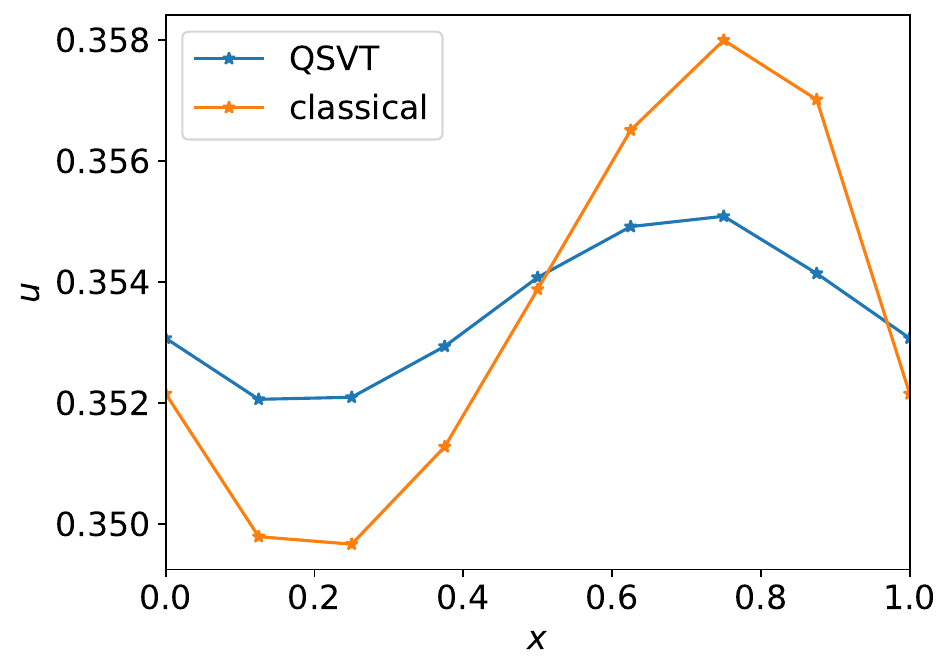} 
        \caption{$e_r=0.0056$}
    \end{subfigure}
\hfill
\begin{subfigure}[b]{0.49\textwidth}
        \centering
        \includegraphics[width=\textwidth]{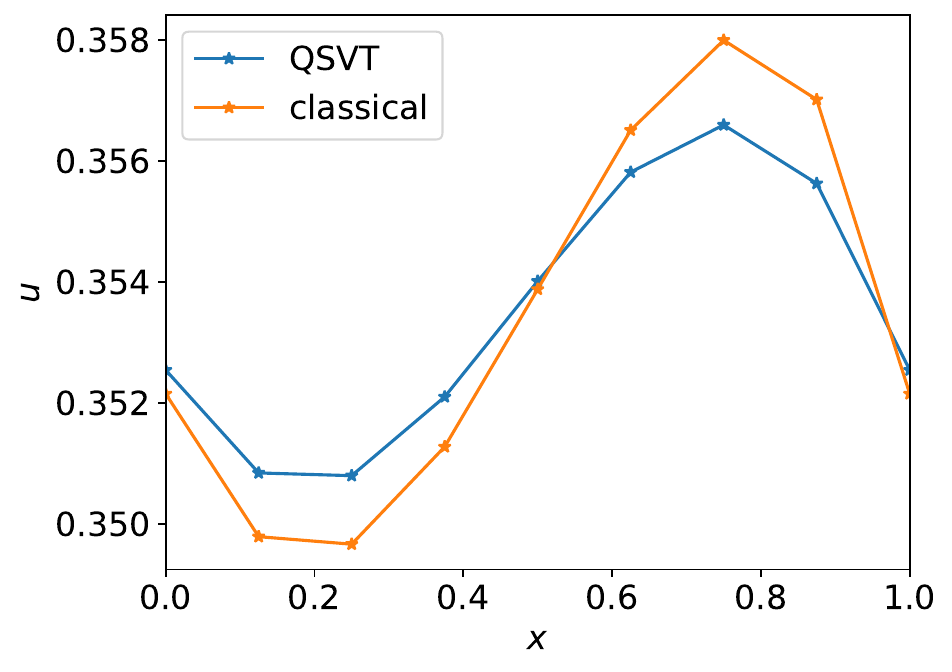} 
        \caption{$e_r=0.0026$}
    \end{subfigure}
    \caption{Comparison between (normalized) classical result and QSVT simulation for the elliptic problem \eqref{eq:ellipticproblem} with $D=1$, $\omega=2$ and $a(x)=1.5+\cos(\omega x)$ (a) for the case $k=3$  and (b) for the case $k=4$. For both cases the relative error in the Euclidean norm $e_r$ has been computed.}
    \label{fig:ellipticprobnum}
\end{figure}

\begin{figure}[htbp]
    \centering
 \begin{subfigure}[b]{0.49\textwidth}
        \centering
        \includegraphics[width=\textwidth]{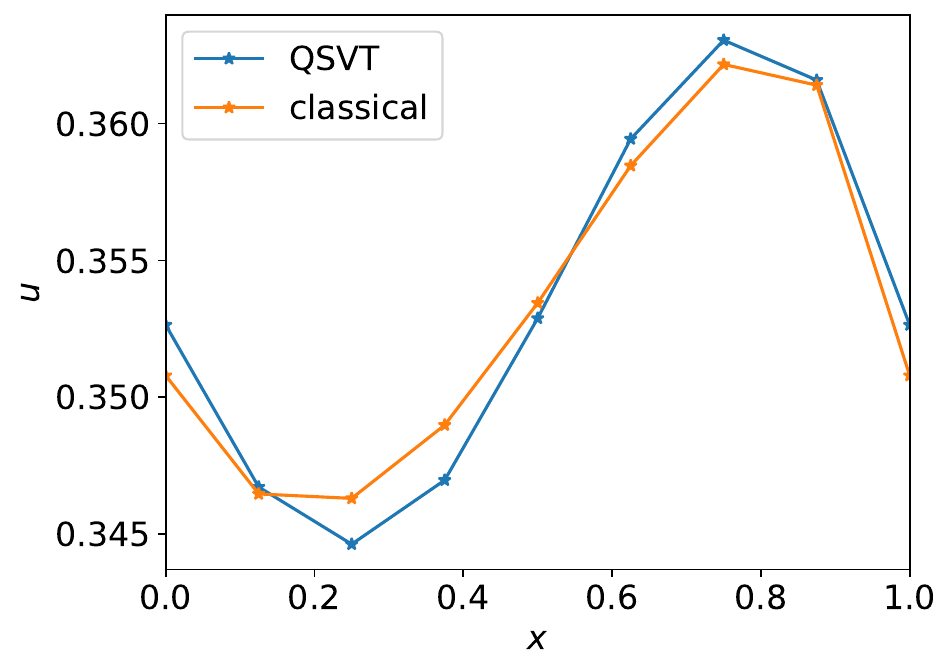} 
        \caption{$e_r=0.0038$}
    \end{subfigure}
\hfill
\begin{subfigure}[b]{0.49\textwidth}
        \centering
        \includegraphics[width=\textwidth]{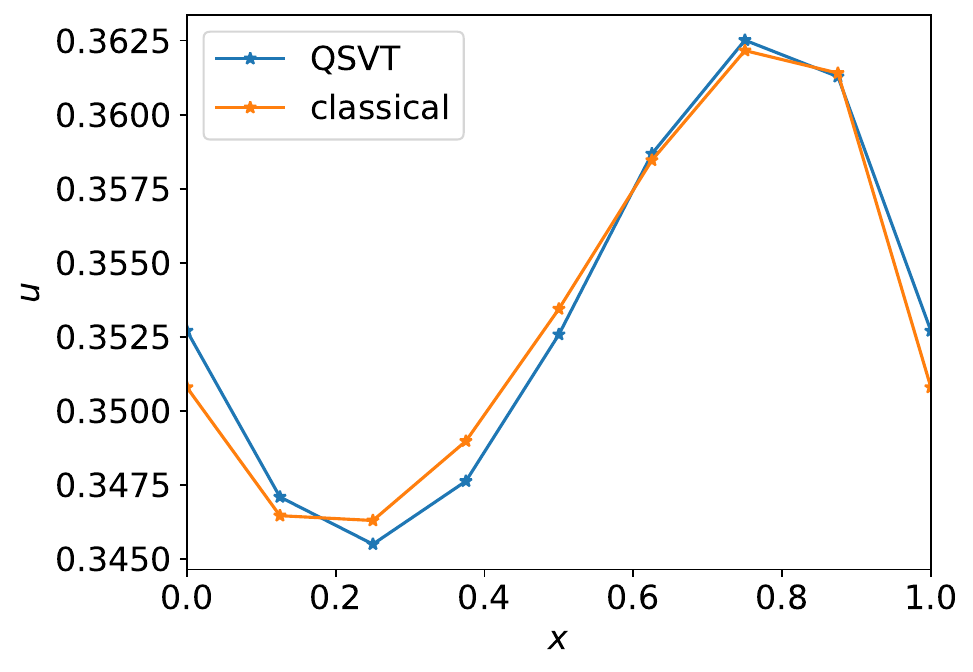} 
        \caption{$e_r=0.0031$}
    \end{subfigure}
    \caption{Comparison between (normalized) classical result and QSVT simulation for the elliptic problem \eqref{eq:ellipticproblem} with $D=0.1$, $\omega=1$ and $a(x)=1.5+\cos(\omega x)$ (a) for the case $k=3$ and (b) for the case $k=4$. For both cases the relative error in the Euclidean norm $e_r$ has been computed.}
    \label{fig:ellipticprobnum2}
\end{figure}

We have numerically verified the proposed methodology using Qiskit and the pyqsp library to implement the QSVT algorithm. The pyqsp library serves as a computationally efficient tool to calculate the corresponding phase factors for a given polynomial transformation of the singular values of the input matrix \cite{PhysRevA.103.042419}.
In Figures \ref{fig:ellipticprobnum} and \ref{fig:ellipticprobnum2} we compare the QSVT result to the ones achieved by a classical linear system solver for the problem \eqref{eq:ellipticproblem} in two different settings. The first case is with $D=1$, $\omega=2$ and $a(x)=1.5+\cos(\omega x)$ in a computational grid using $N=8$ nodes for the periodic $u(0)=u(1)$ solution, and the second case employs $D=0.1$, $\omega=1$. The pyqsp library has been used to find the phase factors for the polynomial approximation of $1/x$ in the interval $[1/k,1]$ using the procedure described in \cite{childs2017quantum} and $k=3$ or 4. The second case achieves a more accurate result with QSVT since the condition number of the matrix is smaller. To achieve higher accuracy and a lower discrepancy with the classical solution, a larger value of $k$ should be used since the condition number of the input matrix is high even for such a coarse spatial discretization.

\begin{figure}[htbp]
    \centering
 \begin{subfigure}[b]{0.49\textwidth}
        \centering
        \includegraphics[width=\textwidth]{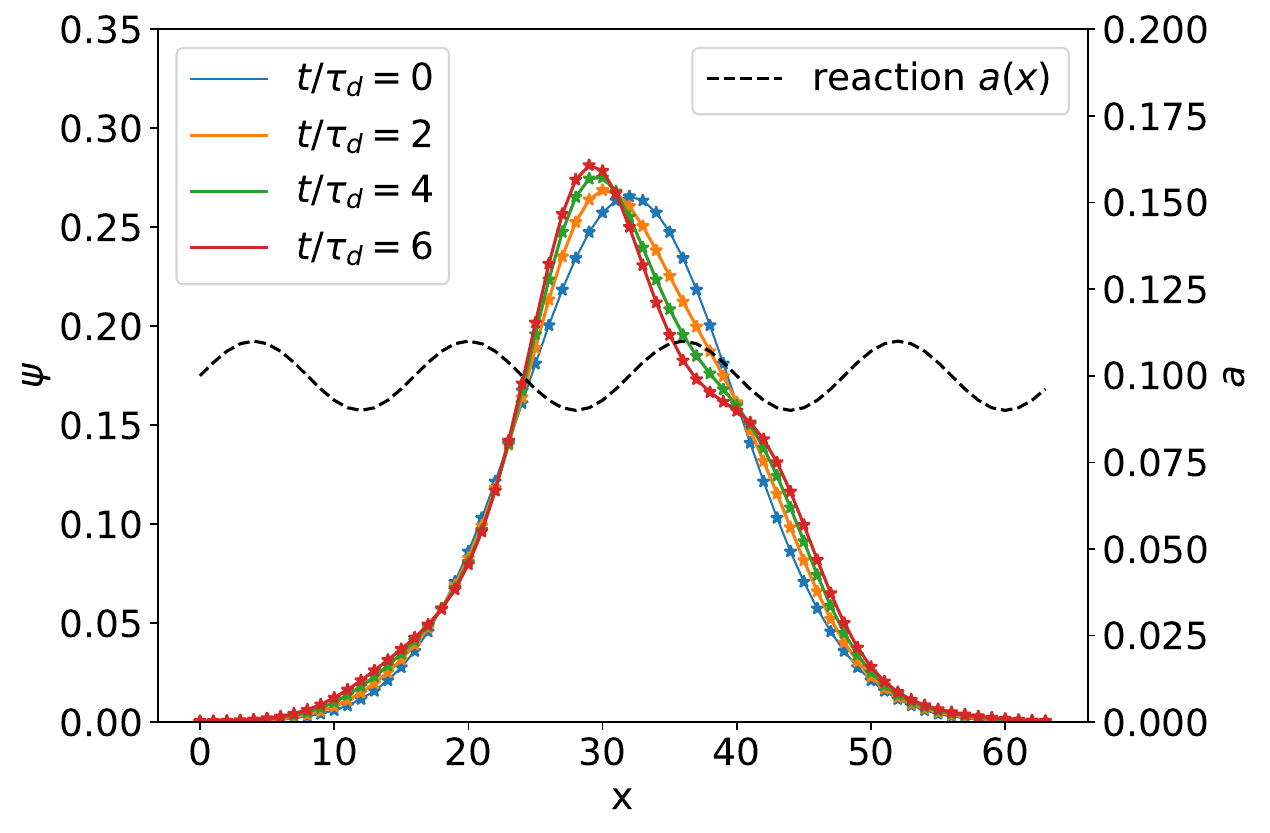} 
    \end{subfigure}
\hfill
\begin{subfigure}[b]{0.49\textwidth}
        \centering
        \includegraphics[width=\textwidth]{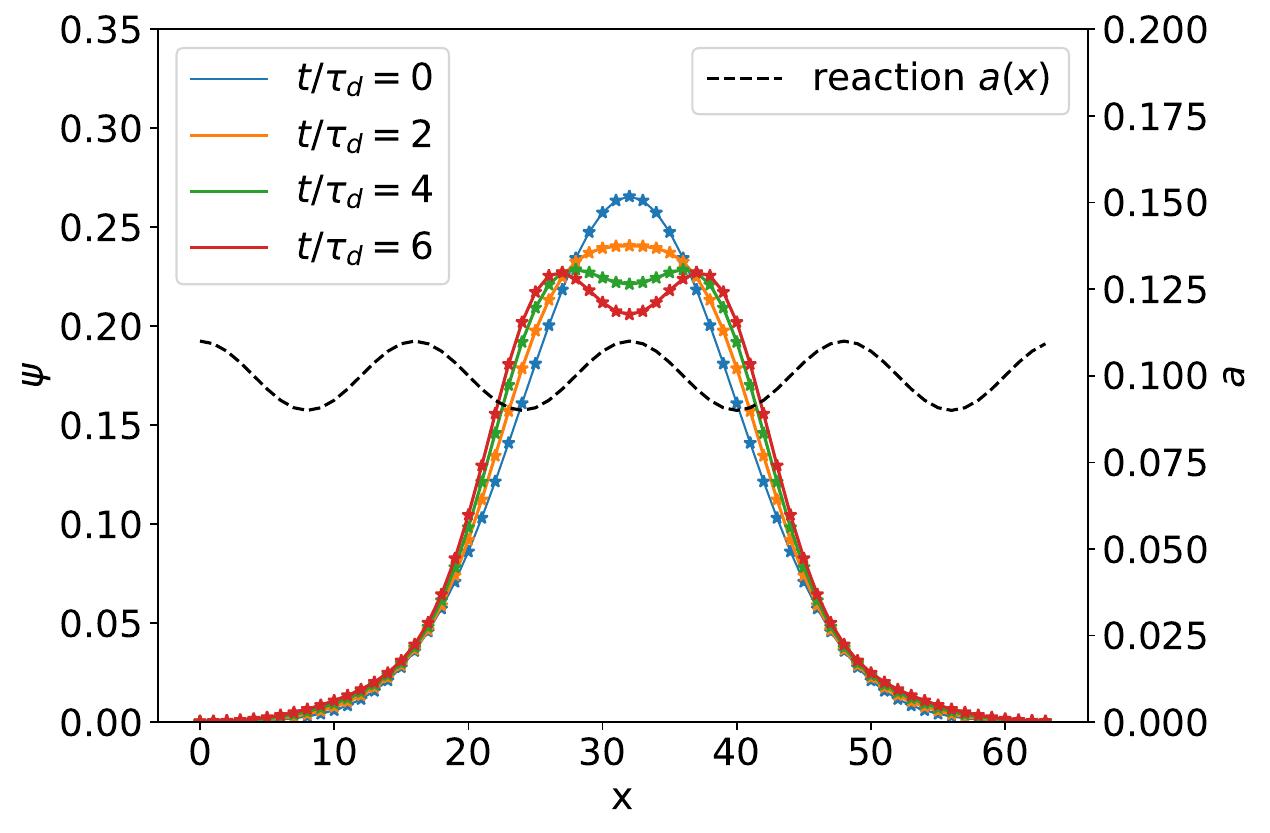} 
    \end{subfigure}
\vspace{0.1cm}
 \begin{subfigure}[b]{0.49\textwidth}
        \centering
        \includegraphics[width=\textwidth]{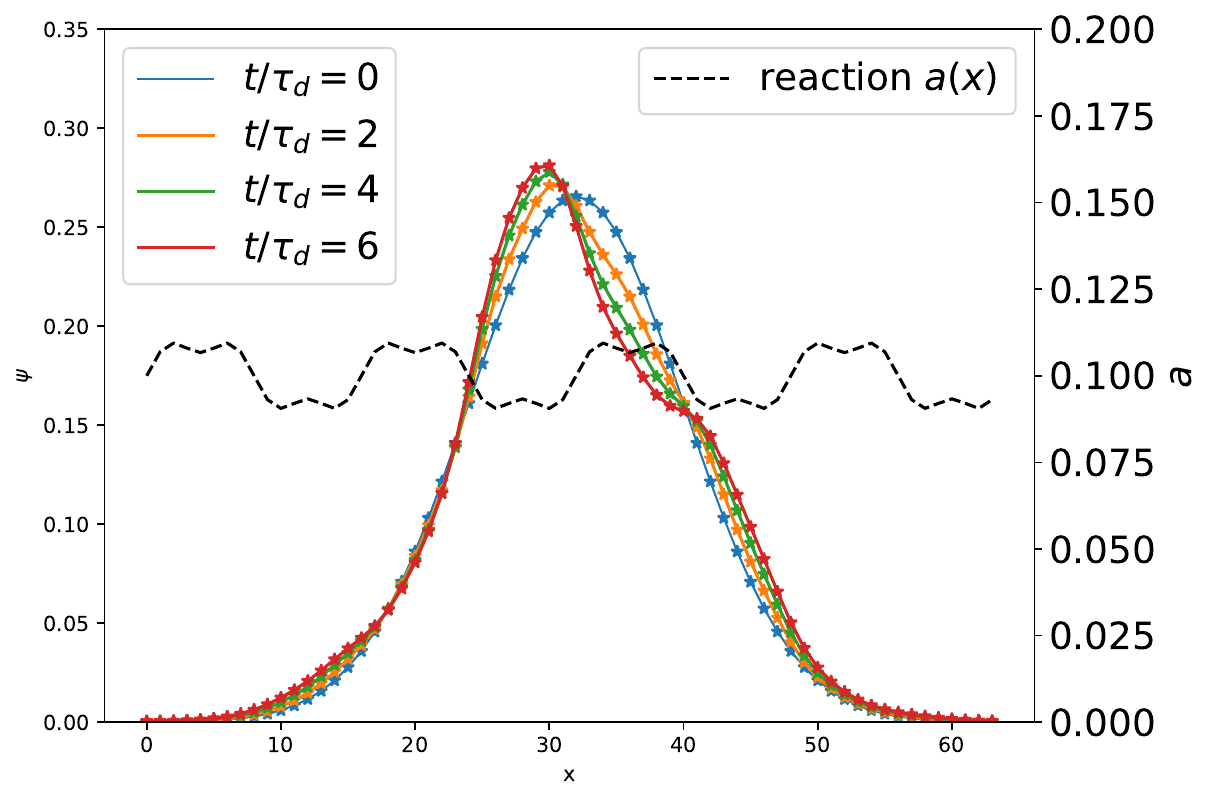} 
    \end{subfigure}
\hfill
\begin{subfigure}[b]{0.49\textwidth}
        \centering
        \includegraphics[width=\textwidth]{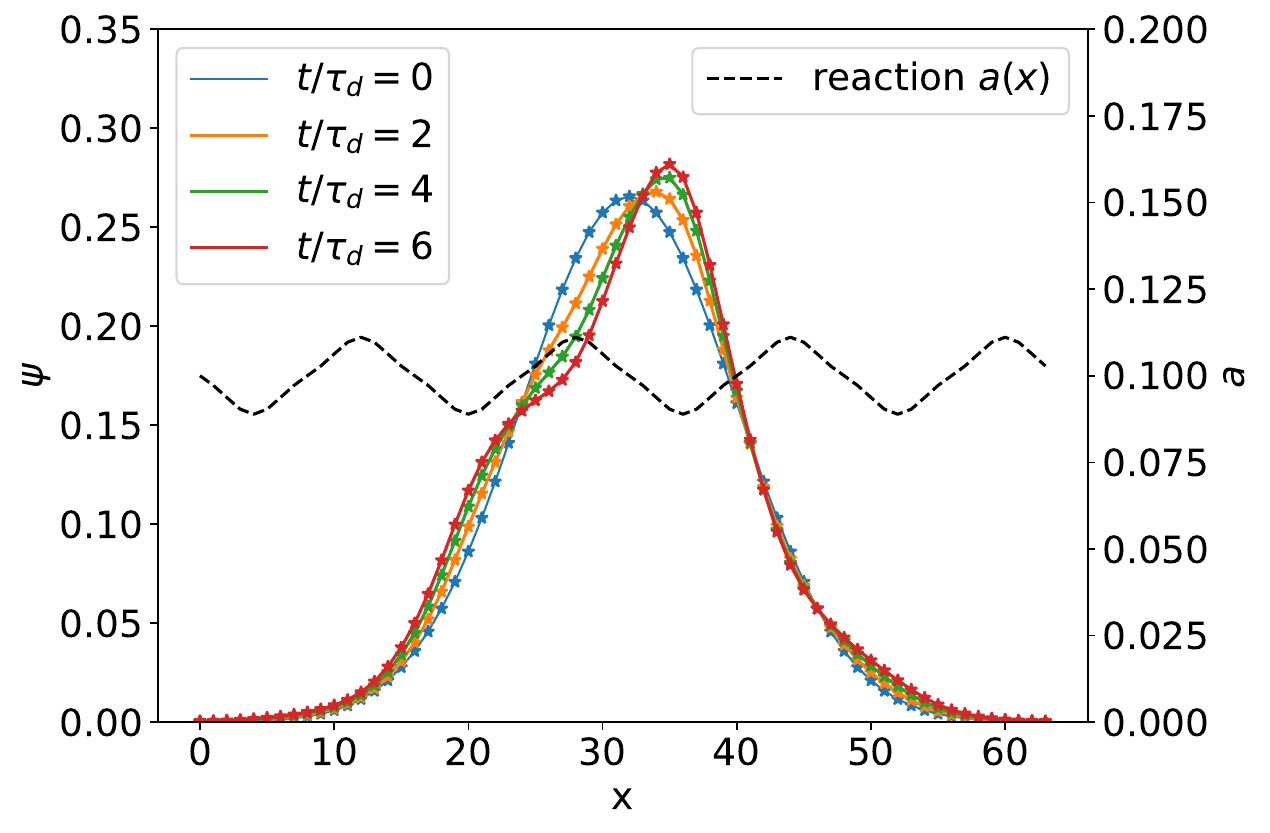} 
    \end{subfigure}
    \caption{QSVT simulation (dots) for a periodically varying reaction term without the advection term $c=0$ implementing a block encoding of $e^{Mt}$ for the matrix $M$ of Equation \eqref{eq:matrixdiscper}, compared with the exact result (solid line) and where the dashed lines represent the reaction term. The four cases shown are with different reaction terms: (a) sine wave, (b) cosine wave, (c) sum of two sines reproducing a square wave and (d) sum of two sines reconstructing a triangular wave.}
    \label{fig:periodicreactionnum}
\end{figure}

For the ADR system we employed again QSVT for the periodically varying reaction term case tackling different configurations for the reaction term. In the simulations we take $c=0$ and a constant value of the diffusion coefficient $D$ in order to have $D/\Delta x^2=0.2$. In this case the pyqsp library has been used to find the phase factors to implement, for a given simulation time $t$, the polynomial approximation of $f(x)=e^{xt}$ as applied to the eigenvalues of the matrix $M$ of Equation \eqref{eq:matrixdiscper}. In Figure \ref{fig:periodicreactionnum} the results of the simulations, for an initial Gaussian profile, are compared to the exact classical simulations computing $e^{Mt}$ for different times expressed in units of the cell diffusion time $\tau_d=\Delta x^2/D$. The state evolves accordingly to the reaction term modifying its shape with time as expected. Four cases are shown, in the upper row (a) with a reaction term of $a(x)=0.1+0.01\sin(\frac{2\pi}{16}x)$ and (b) with $a(x)=0.1+0.01\cos(\frac{2\pi}{16}x)$ (b). In the lower row (c) with reaction of $a(x)=0.1+0.01\sin(\frac{2\pi}{16}x)+\frac{0.01}{3}\sin(\frac{6\pi}{16}x)$ and (d) $a(x)=0.1-0.01\sin(\frac{2\pi}{16}x)+\frac{0.01}{9}\sin(\frac{6\pi}{16}x)$.


\section{Conclusion}\label{sec:conclusion}
In this work we presented a low volume (circuit depth and size) quantum circuit for block encoding sparse matrices with an inherent structure. In particular, we exploit the underlying periodicity to achieve better efficiency than general-purpose methods, a consistent finding in the block encoding research \cite{Kuklinski2025EfficientBR}. For the case of a matrix with a structure on the main diagonal the circuit achieves a gate complexity that scales linearly with the number of qubits $\mathcal{O}(n)$. The proposed methodology is also capable of handling signals consisting of different frequency components, making it versatile for a wide range of applications.
Our work can be considered as a generalization of the GQSP method where the frequency $\omega$ is not restricted to integer multiples of $2\pi/N$, enabling an efficient way to express generic signals \cite{PRXQuantum.5.020368}.
The advantage of the proposed methodology compared to the existing literature lies in exploiting appropriately periodicity to reduce depth and avoid generic sparse or dense encodings. Our approach is particularly advantageous when the signal to be decomposed admits a compact Fourier representation so that only a small number of terms for the LCU is required. We have demonstrated the utility of the proposed block encoding through various use cases in simulating classical dynamics in one spatial dimension.

Despite the positive findings, several points still need to be addressed to further develop the proposed block encoding approach. For a generic signal to be encoded in a diagonal non-unitary matrix there is the possibility of a more efficient quantum circuit. While the LCU method for handling multiple frequencies requires ancilla qubits, which may limit scalability, alternative approaches may offer reduced resource overhead \cite{vasconcelos2025methodsreducingancillaoverheadblock}. Moreover, the sub-normalization factor in the LCU setting, here considered, may be sub-optimal.
A key open problem is the development of a more general framework for frequency selection, potentially based on matrix-access oracle models.

In summary, our work provides a practical framework for efficiently block encoding structured sparse matrices on quantum hardware, offering a starting point for simulating classical and quantum systems with inherent periodicity on quantum computers.

\bibliographystyle{unsrturl}
\bibliography{bibliography}

\end{document}